\documentclass[11pt]{article}

\date{20 October 2025}

\usepackage{sty/jn-standard}
\usepackage{sty/inclusion}
\usepackage{sty/macros}
\usepackage{sty/specificMacros}
\usepackage{mathpartir}

\begin{document}

\author{Jacob Neumann}
\affil{Reykjavik University\\ Menntavegur 1, 102 Reykjavik, Iceland\\
\texttt{jacobn@ru.is}}

\title{A Judgmental Construction of Directed Type Theory}

\maketitle
\begin{abstract}
We reformulate recent advances in directed type theory---a type theory where the types have the structure of synthetic (higher) categories---as a logical calculus with multiple context \enquote{zones}, following the example of Pfenning and Davies. This allows us to have two kinds of variables---\enquote{neutral} and \enquote{polar}---with different functoriality requirements. We focus on the lowest-dimension version of this theory (where types are synthetic \textit{preorders}) and apply the logical language to articulate concepts from the theory of rewriting. We also take the occasion to develop the categorical semantics of dual-context systems, proposing a notion of \textit{dual CwF} to serve as a common structural base for the model theories of such logics.

\end{abstract}


\section{Introduction}\label{intro}
    %% ./intro.tex
%%
%% By Jacob Neumann (jacobneu.com)
%% October 2025
%%

\subsection{Directed type theory}
A central notion in modern type theories, particularly dependent type theories in the tradition of Martin-L{\"o}f~\cite{MLTT1}, is that of \textit{identity types}. We can understand the purpose of identity types as \textit{internalizing metatheoretic equality}, that is, taking the statement that two terms $s$ and $t$ (of the same type) are \textit{equal}---which is traditionally a \textit{judgment}, an assertion made \textit{about} the objects of the language but not \textit{in} the language itself---and making it a construct within the language, the type $\Id(s,t)$. In a language equipped with this feature, it can be proved \textit{within} the language that $s$ is equal to $t$, by constructing a piece of data, a term of type $\Id(s,t)$.\footnote{This is particularly relevant, for instance, in the field of \textit{interactive theorem proving}, where the goal is to represent mathematical proofs within a programming language. Identity types are a natural addition to such a language, as they represent the equality of two things; proving that things are \textit{equal} to each other is arguably the fundamental kind of proof mathematicians perform.} The nuances, pitfalls, and possibilities of including identity types in a language have occupied the attention of type theorists for decades, and will no doubt continue to be a topic of intense investigation.

However, type theorists have only recently begun to grapple with \keyword{directed identity types} (also known as \textit{hom-types}), the analogous internalization of \textit{reduction/rewriting}. The informal idea of the directed identity type $\hom(s,t)$ is that it's supposed to be \textit{like an identity type, but with a direction}. More exactly, we desire hom-types to be reflexive (the type $\hom(t,t)$ ought always to be inhabited) and transitive (we ought to be able to combine an element of $\hom(s,t)$ with an element of $\hom(t,u)$ to obtain an element of $\hom(s,u)$) but \textit{not symmetric}: unlike identity types (which usually \textit{are} provably symmetric), we don't want a term of $\hom(s,t)$ to give us a term of $\hom(t,s)$ in general. This reflects the fact that metatheoretic reduction generally goes one way: $s$ might reduce to $t$ but not vice-versa. Soundly designing a type theoretic language that behaves this way---a \keyword{directed type theory}---has proved to be a surprisingly challenging task, for a variety of technical reasons.

\subsection{Judgmental directed type theory}
Numerous approaches to directed type theory have been put forward in the literature. Were one to arrange these theories into some kind of taxonomy, the top-level division would surely be the \textit{manner of introducing the directed identity types}. The prevailing directed type theory of the past several years has been the \textit{simplicial type theory} originally proposed by Riehl and Shulman~\cite{riehl2017type}, further expanded by Weinberger~\cite{weinberger2022asynthetic}, and recently refined by Gratzer et al.~\cite{gratzer2024directedunivalencesimplicialhomotopy,gratzer2025yonedaembeddingsimplicialtype}. The simplicial approach to directed type theory is to first axiomatize a directed interval type $\mb I$ and \textit{define} directed identity in terms of it:
\begin{align*}
\hom[A](s,t) &:= \sum_{f\colon \mb I \to A} \Id(f\;0,s)\times\Id(f\;1,t)
\tag*{{}\cite[Definition 2.2]{gratzer2024directedunivalencesimplicialhomotopy}.}
\end{align*}
This approach comes with numerous benefits, such as superior computational properties.\footnote{Simplicial type theory defines hom-types in a manner reminiscent of how \textit{cubical type theory}~\cite{bezem2013amodel,cohen2018cubical} defines identity types, and thereby enjoys many of the computational advantages of the latter.} However, the semantics of simplicial theory are rather complex---they must be at least as complex as homotopy type theory, of which simplicial type theory is an extension.

Also, if we wish to explore directed identity as a reflection of metatheoretic reduction, simplicial type theory is much more than we need. Simplicial type theory is intended as a synthetic language for $\infty$-categories: we can understand the hom-types in a directed type theory as endowing a type with a \keyword{synthetic category structure}: the terms of a type $A$ represent the objects of the category, the terms of type $\hom(s,t)$ represent morphisms from $s$ to $t$, reflexivity corresponds to identity morphisms, etc. However, since $\hom(s,t)$ is \textit{also} a type, we can iterate this process, considering hom-terms \textit{between} hom-terms, and hom-terms between those hom-terms, and so on forever. Simplicial type theory is designed to allow its users to manipulate this $\infty$-categorical structure effectively, and the research into it is geared towards that end. However, rewriting systems are not $\infty$-categories, they are preorders, i.e. 0-categories. Therefore, we can get away with much simpler systems with more straightforward semantics.

The standard approach in \textit{un}directed type theory---dating back to Martin-L{\"o}f~\cite{MLTT1}---is to introduce identity types as a primitive type-former, governed by several inference rules. In particular, the elimination principle for identity types, the \textbf{J-rule}, is asserted along with a $\beta$-law which holds up to \textit{judgmental equality}.\footnote{As opposed to, say, cubical type theory, where there J-rule is merely a theorem, and its $\beta$-law holds only \textit{propositionally}.} We shall use the term \textbf{judgmental directed type theory} for a theory which does likewise for directed identity: articulate hom-types as a primitive type-former, along with a directed J-rule whose $\beta$-law holds judgmentally.\footnote{This is not to suggest that other directed type theories do not observe Martin-L{\"o}f's distinction between judgment and proposition, but simply that they do not articulate their directed identity types judgmentally.} Our purpose in the present work is to articulate a judgmental directed type theory.

Our approach to directed type theory will reconcile the type theories of North~\cite{north2019} and Neumann-Altenkirch~\cite{neumann2025asynthetic,neumann2025ageneralized}.
One benefit of these theories is that their semantics are very straightforward: they are interpreted in 1-categories (namely the \textit{category model}, a directed analogue of Hofmann and Streicher's \textit{groupoid model}~\cite{groupoidModel}). This is in contrast to simplicial type theory (and similar theories, e.g. \cite{weaverLicata}), as well as the \textit{virtual equipment type theory (VETT)} of New and Licata~\cite{new2023formal} (which is a \enquote{judgmental directed type theory} in the sense given above); the semantics of these theories involve extraordinarily complex mathematical constructions, which are relatively removed from ordinary 1-categories.

One reason we care about semantic complexity is that a straightforward semantics makes it easier to perform metatheoretic arguments, e.g. independence arguments---demonstrating what \textit{cannot} be constructed within the theory. This is of vital importance to directed type theory: if we are to verify that a putative directed type theory is indeed directed, we must be able to argue that symmetry is not provable in general, that $p\colon\hom(s,t)$ cannot always be turned around to $p\inv\colon\hom(t,s)$. This concern is especially acute for judgmental directed type theories: in \textit{un}directed type theory, the same J-rule is used to prove both the symmetry and transitivity of equality; why should a directed J-rule be able to prove the latter but not the former? Unfortunately, New and Licata neglect this question, and it's not obvious why their $\mathrm{ind}_\to$ term-former cannot be used to prove that all the categories in VETT's formal category theory aren't automatically groupoids. In the Neumann-Altenkirch theory, on the other hand, it's easy to conduct an independence proof~\cite[Theorem 3.1.20]{neumann2025ageneralized} modeled on the Hofmann-Streicher proof~\cite[Theorem 5.1]{groupoidModel} of the independence of the \textit{uniqueness of identity proofs} principle (UIP) from the rules of Martin-L{\"o}f type theory; this proof will work equally well in our system.

Where North and Neumann-Altenkirch differ is the precise way they restrict the J-rule to prevent symmetry. Both theories follow Licata and Harper~\cite{licataHarper} in imposing a \textit{polarity calculus}: a modal typing discipline which annotates terms as either \textit{co-} or \textit{contra-variant} (\enQuote{positive} or \enQuote{negative}, respectively). By requiring the domain term $s$ to be annotated negative and the codomain term $t$ annotated positive in order for $\hom(s,t)$ to be well-formed, the proof of symmetry can be prevented but the proof of transitivity undisturbed. The issue comes with reflexivity: how is the judgment $\refl_s\colon\hom(s,s)$ supposed to be well-formed under this polarity calculus, if $s$ must appear both positively and negatively? North restricts to polarity-invariant \textit{types}, requiring $s$ to be invariant in order for $\refl$ (and consequently, J) to be given; Neumann-Altenkirch, on the other hand, restrict to polarity-invariant \textit{contexts}, wherein any term can be freely coerced between \enquote{polarities}. We present the \enQuote{least upper bound} of these two theories, generalizing both approaches into a common J-rule (which still cannot prove symmetry!).

\subsection{Dual-context theories}
Logicians have developed several logical calculi that employ multiple \textit{context zones} to  contain different kinds of hypotheses, i.e. contexts of the form $\Gamma\mid\Delta$, where $\Gamma$ and $\Delta$ are the \enquote{zones}, with different rules for their variables. For instance, Girard~\cite{girard1993unity} employed zoned contexts to encompass classical and intuitionistic sequents in the same logic; Pfenning and Davies~\cite{pfenning2001judgmental} expounded a logic with a context zone for hypothesizing that a proposition is \textit{valid} and one for hypothesizing a proposition is \textit{true}; and, more recently, Shulman~\cite{shulman2018brouwer} gave a type theory for synthetic topological constructions with \enQuote{crisp} and \enQuote{cohesive} context zones, such that a construction depending on several variables is continuous with respect to the synthetic topology of the cohesive variables, but possibly discontinuous with respect to the crisp variables. See e.g. \cite[Introduction]{kavvos2020dual} for a more thorough overview of this area of research.

We employ this technology towards judgmental directed type theory. Our types have the structure of synthetic categories, and any construction made within the theory is automatically \textit{functorial} with respect to the category structure of the context. The issue that North and Neumann-Altenkirch confronted is that identity morphisms ($\refl$) cannot be functorial with respect to both endpoints, as they have opposite variances. This is solved by \textit{neutralizing} the polarity, i.e. only requiring $\refl$ to be functorial with respect to a groupoid context. This is the function of dual-zoned contexts for directed type theory: for the context $\Gamma\mid\Delta$, the zone $\Gamma$ will be structured as a synthetic \textit{groupoid}, so the difference between co- and contra-variance with respect to variables from $\Gamma$ disappears. So, we'll find, terms in context $\Gamma\mid\bullet$, i.e. where the \enQuote{polarized zone} $\Delta$ is empty, will be able to overcome the restrictions of variance to the extent we need.

We also take the opportunity to develop the semantics of this theory. In addition to updating North's semantics for directed type theory---the \textit{category model}---to match the dual-context situation, we consider how to define an abstract notion of \enquote{model} for dual-context type theories. To our knowledge, there is no general-purpose model theory for dual-context type theories, analogous to how Dybjer's \textit{categories with families} (or the many equivalent notions) serve as a common structural basis for type theories. We propose such a notion, upon which (we conjecture) model notions for the various dual-context type theories can be developed.

\section{Syntax of Dual-Context Directed Type Theory}\label{typeTheory}
    %% ./typeTheory.tex
%%
%% By Jacob Neumann (jacobneu.com)
%% October 2025
%%

In this section, we develop the syntax of the theory. As mentioned, our contexts will be of the form $\Gamma\mid\Delta$, where we'll refer to $\Gamma$ as the \textit{neutral zone} and $\Delta$ as the \textit{polar(ized) zone}. We follow the same convention as \cite{shulman2018brouwer}: the neutral zone will consist of judgments of the form $x::A$, and the polar zone of judgments $x\colon A$. For this section, we'll use named variables (omitting the usual side-conditions of freshness, avoidance of capture, etc.), but in the semantics we'll make this more precise with the use of de Bruijn indices.

\begin{figure}
\begin{mathpar}
\inferrule[Empty]{\ }{\Empty\mid\bullet\isCtx}
\and \inferrule[Extend-Polar]{\Gamma\mid\Delta\vdash A\isType}{\Gamma\mid\Delta,x: A\isCtx}
\and \inferrule[Extend-Neutral]{\Gamma\mid\bullet\vdash A\isType}{\Gamma,x::A\mid\bullet\isCtx}
\\
\inferrule[Type Negation]{\Gamma\mid\Delta\vdash A\isType}{\Gamma\mid\Delta\vdash A^-\isType}
\and\inferrule[Negation Involution]{\Gamma\mid\Delta\vdash A\isType}{\Gamma\mid\Delta\vdash (A^-)^- \jEqual A}
\end{mathpar}
\caption{Basic rules of the polarity calculus }\label{tag:polarityCalcRules}

\end{figure}

We have the usual judgments of type theory:
\begin{alignat*}{2}
&\Gamma\mid\Delta\isCtx \qquad&&\text{\enquote{$\Gamma\mid\Delta$ is a well-formed context}}\\
&\Gamma\mid\Delta\vdash A\isType \qquad&&\text{\enquote{$A$ is a well-formed type in context $\Gamma\mid\Delta$}}\\
&\Gamma\mid\Delta\vdash t\colon A \qquad&&\text{\enquote{$t$ is a term of type $A$ in context $\Gamma\mid\Delta$}}\\
&\Gamma\mid\Delta\vdash A\jEqual A' \qquad&&\text{\enquote{$A$ and $A'$ are judgmentally-equal types in context $\Gamma\mid\Delta$}}\\
&\Gamma\mid\Delta\vdash t\jEqual t' \qquad&&\text{\enquote{$t$ and $t'$ are judgmentally-equal terms in context $\Gamma\mid\Delta$}}
\end{alignat*}
The basic rules for manipulating contexts are given in \cref{tag:polarityCalcRules}: both context zones are finite lists of typed variable declarations, where we write $\bullet$ for the empty list (as usual, we omit the leftmost $\bullet$ for nonempty lists). Since this is a directed type theory, all types come equipped with a synthetic category structure, and by extension so do the contexts. In a judgment $\Gamma\mid\Delta\vdash\mc J$, the polarized context zone $\Delta$ depends functorially on the neutral context zone $\Gamma$, and the type or term asserted by $\mc J$ depends functorially on both. However, the neutral context zone $\Gamma$ will be carefully maintained as a synthetic \textit{groupoid}, so the distinction between co- and contra-variance is obliterated with respect to $\Gamma$. $\Delta$, on the other hand, is allowed to be genuinely \textit{directed}, so the functorial dependence of $\mc J$ on $\Delta$ must observe co- and contra-variance---this is the sense in which variables from $\Delta$ are \enQuote{polarized}. For this reason, the neutral zone $\Gamma$ cannot be extended by a type $A$ which contains polarized variables---hence $\Delta$ must be empty in the premise of \textsc{Extend-Neutral}.

To indicate the distinction between covariant and contravariant functors, category theorists employ the notion of \textit{opposite categories}, e.g. writing $F\colon\mc C\op\to\mc D$. We, following \cite{north2019}, do likewise: for every type $A$, there is a corresponding type $A^-$ (in the same context), its \textit{opposite}. Later, once we introduce hom-types, we'll be able to express that $A^-$ has the opposite synthetic category structure as $A$; as \cite[Chapter 2]{neumann2025ageneralized} observes, synthetic category theory (in the form of judgmental directed type theory) must develop a theory of duality \textit{before} articulating the notion of \enquote{synthetic category}, contrary to the usual (analytic) category theory, which must define \enquote{category} in order to speak of \enquote{opposite categories}. We assert judgmentally that this type-negation operation is involutive. For the sake of brevity, we omit congruence laws (e.g. that if $A\jEqual B$, then $A^-\jEqual B^-$) and the stability of negation under substitution (i.e. that $A^-[t/x]\jEqual (A[t/x])^-$).

\begin{figure}
\begin{mathpar}
\inferrule[Core Type]{\Gamma\mid\Delta\vdash A\isType}{\Gamma\mid\Delta\vdash A^\flat \isType}
\and \inferrule[Core Variable]{\ }{\Gamma,y::A,\Gamma'\mid\Delta\vdash y\colon A^\flat}
\and \inferrule[Core-Negation]{\Gamma\mid\Delta\vdash A\isType}{\Gamma\mid\Delta\vdash (A^-)^\flat \jEqual (A^\flat)^-}
\and \inferrule[Core-Idem]{\Gamma\mid\Delta\vdash A\isType}{\Gamma\mid\Delta\vdash (A^\flat)^\flat \jEqual A^\flat}
\\
\inferrule[Core Intro+]{\Gamma\mid\bullet\vdash t\colon A}{\Gamma\mid\Delta\vdash \pair{+\flat}t\colon A^\flat}
\and \inferrule[Core Intro-]{\Gamma\mid\bullet\vdash s\colon A^-}{\Gamma\mid\Delta\vdash \pair{-\flat}s\colon A^\flat}
\and \inferrule[Core Elim+]{\Gamma\mid\Delta\vdash e\colon A^\flat}{\Gamma\mid\Delta\vdash \pair{+}e\colon A}
\and \inferrule[Core Elim-]{\Gamma\mid\Delta\vdash e\colon A^\flat}{\Gamma\mid\Delta\vdash \pair{-}e\colon A^-}
\\
\inferrule[Core $\beta+$]{\Gamma\mid\bullet\vdash t\colon A}{\Gamma\mid\Delta\vdash \pair{+}\pair{+\flat}t \jEqual t}
\and \inferrule[Core $\beta-$]{\Gamma\mid\bullet\vdash s\colon A^-}{\Gamma\mid\Delta\vdash \pair{-}\pair{-\flat}s \jEqual s}
\and \inferrule[Core $\eta+$]{\Gamma\mid\bullet\vdash e\colon A^\flat}{\Gamma\mid\bullet\vdash \pair{+\flat}\pair{+}e \equiv e}
\and \inferrule[Core $\eta-$]{\Gamma\mid\bullet\vdash e\colon A^\flat}{\Gamma\mid\bullet\vdash \pair{-\flat}\pair{-}e \equiv e}
\and \inferrule[Core Subst]{\Gamma,x::A\mid\Delta\vdash m\colon M \\ \Gamma\mid\Delta\vdash e\colon A^\flat}{\Gamma\mid\Delta\vdash m[e/x]\colon M[e/x]}
\end{mathpar}
\caption{Rules for core types. }\label{tag:coreTypeRules}

\end{figure}

In \cref{tag:coreTypeRules} we specify \textit{core types} and their interaction with the neutral context zone; this is the connective tissue between the type theories of North and Neumann-Altenkirch. North makes use of core types (which we write as $A^\flat$ instead of $A^{\textsf{core}}$) to address the issue of $\refl$'s divariance: a term $e\colon A^\flat$ can be coerced to either $\pair+ e\colon A$ or $\pair- e\colon A^-$ (which she denotes $\mathsf{i}\;e$ and $\mathsf{i}^\mathsf{op}\;e$, respectively), thereby allowing \enQuote{the same term} to appear both in the positive and negative position. Neumann and Altenkirch, objecting to the apparent restriction of $\refl$ and $\J$ to \textit{only} being anchored at core terms (which North does not provide a means for constructing), avoid the use of core types: they observe that, in a neutral context (in our system, a context of the form $\Gamma\mid\bullet$), it is possible to coerce between the polarities, sending $s\colon A^-$ to $-s\colon A$ and vice-versa. But this isn't as removed from North's approach as they seem to think: in a neutral context, we can \textit{also} coerce $s\colon A^-$ to a term of type $A^\flat$, which we write $\pair{-\flat}s$ (and likewise $\pair{+\flat}t\colon A^\flat$ for $t\colon A$). Indeed, semantically $\pair{-\flat}s$ just \enQuote{packages} the negative term $s$ with Neumann-Altenkirch's positive term $-s$, forming a single, neutral term. In this way, the Neumann-Altenkirch coercion operations can be defined in terms of the North ones:
\begin{align}
-s := \pair{+}\pair{-\flat}s \qquad\qquad -t := \pair{-}\pair{+\flat}t.
\label{tag:termCoerceDefn}
\end{align}
We'll see below that this allows us to view both treatments of $\refl$ and $\J$ as instances of a single, common one.
The rules \textsc{Core Variable} and \textsc{Core Subst} show us that a variable $x::A$ in the neutral zone is essentially a variable of $A^\flat$---indeed, in the category model semantics these will be precisely the same.

We have now prepared the ground for a development of directed type theory; the rules for hom-types are given in \cref{tag:homTypeRules}. As in North, the hom-type can be formed in an arbitrary context $\Gamma\mid\Delta$; $\hom$ has its domain term annotated negative/contravariant and its codomain term annotated positive/covariant. Our \textsc{Hom Intro} rule for $\refl$ also matches North's exactly, but generalizes the Neumann-Altenkirch rule (for each $s\colon A^-$ in a neutral context, obtain $\refl_s\colon\hom(s,-s)$) by way of \cref{tag:termCoerceDefn} and \textsc{Core }$\beta-$.

Both North and Neumann each have \textit{two} elimination principles for their hom-types:\footnote{Neumann-Altenkirch only includes only the \enQuote{forward}/\enQuote{positive}/\enQuote{coslice} version.} one rule based at the domain and inducting on hom-terms \textit{out} of it (\enQuote{\textsc{right} \textsf{hom} \textsc{elimination}}, $\mathsf{e}_\mathsf R$, for North; \enQuote{coslice path induction}, $\J[+]$, for Neumann); and, dually, one based at the codomain and inducting on hom-terms \textit{into} it (\enQuote{\textsc{left} \textsf{hom} \textsc{elimination}}, \enQuote{slice path induction}). Our rule encapsulates both of Neumann's rules (included here as \cref{tag:oneSidedJRules} for reference) in a single one---hence the notation $\J[\pm]$---by simultaneously inducting in both directions: in order to prove a claim $M(u,v)$ depending on a hom-term $u$ \textit{into} $y$ and a hom-term $v$ \textit{out of} $y$, it suffices to supply a proof of $M(\refl_y,\refl_y)$. Neumann's $\J[+]$ (respectively, $\J[-]$) operator can be obtained by a type family $M(u,v)$ which doesn't actually depend on $u$ (resp., which doesn't depend on $v$) and substituting in $\refl$ for $u$ (resp. $v$) in the final result. The $\beta$ laws obtained by this specialization will then match his $\beta$ laws as well. North's rules are slightly more elaborate (they contain an extra \enQuote{layer} of type families), but this level of generality is not needed for any of the constructions North does (and, in any case, can again be obtained from our rule in the presence of $\Pi$-types).

\begin{figure}
\begin{mathpar}
\inferrule[Hom Form]{\Gamma\mid\Delta\vdash A\isType}{\Gamma\mid\Delta,x\colon A^-,z\colon A\vdash \hom[A](x,z)\isType}
\and \inferrule[Hom Intro]{\Gamma\mid\Delta\vdash e\colon A^\flat}{\Gamma\mid\bullet\vdash \refl_e\colon\hom(\pair- e,\pair+ e)}
\and \inferrule[Hom Elim$\pm$]{
    \Gamma,y::A\mid x\colon A^-, z\colon A, u\colon\hom(x,\pair+ y), v\colon\hom(\pair- y, z)\vdash M(u,v)\isType \\
    \Gamma,y::A\mid\bullet \vdash m\colon M(\refl_y,\refl_y)
}
{\Gamma,y::A\mid x\colon A^-, z\colon A, u\colon\hom(x,\pair+ y), v\colon\hom(\pair- y, z)\vdash \J[\pm]\;m\colon M(u,v)}
\and \inferrule[Hom $\beta\pm$]{
    \Gamma,y::A\mid x\colon A^-, z\colon A, u\colon\hom(x,\pair+ y), v\colon\hom(\pair- y, z)\vdash M(u,v)\isType \\
    \Gamma,y::A\mid\bullet \vdash m\colon M(\refl_y,\refl_y)
}
{\Gamma,y::A\mid \bullet\vdash (\J[\pm]\;m)[\refl_y/u,\refl_y/v] \jEqual m}

\end{mathpar}
\caption{Rules for hom-types. }\label{tag:homTypeRules}

\end{figure}

This bidirectional J-rule allows us to make a cleaner and more principled definition of \textit{composition} of hom-terms (that is, the transitivity of directed equality) than possible in our predecessors:\footnote{North~\cite[p. 4]{north2019} only briefly addresses the possibility of \enQuote{left transportation} and the possibility of using it to define composition, but the rigidity of working in arbitrary polarized contexts makes it such that it usually doesn't make sense to contemplate whether the two notions of \enQuote{composition} coincide (and she doesn't address the question in any event). Neumann-Altenkirch, as mentioned, does not contemplate the backwards form of path induction. Neumann~\cite[Subsection 3.2.1]{neumann2025ageneralized} does define composition both ways, but has to make a rather \textit{ad-hoc} axiom to assert their coincidence.}
\mkDefn{syntheticComposition}
The downside of this being \enQuote{unbiased} to either the left or right (to slices or coslices) is that we do not get a judgmental computation rule in either direction, i.e. saying $f\cdot\refl\jEqual f$ or $\refl\cdot g\jEqual g$. Semantically, both equations are validated by our intended model, so we would be justified in just baldly asserting them (this is essentially what Neumann does), but we'll satisfy ourselves with being able to prove them up to propositional equality (see below).

Let us repeat here the reasoning from Neumann-Altenkirch as to \textit{why symmetry is not provable from this J-rule}. As we indicated in the introduction, we count this as a fundamental issue for directed type theories: a \enQuote{directed} type theory where symmetry is provable in general is not a \textit{directed} type theory. The informal \enQuote{proof} of putting $\refl\inv := \refl$ and then concluding \enQuote{by induction} that there is an operation inverting any $p\colon\hom(s,t)$ to $p\inv\colon\hom(-t,-s)$ is not a valid application of $\J[\pm]$. This is where the \enQuote{polarity calculus} we've established plays its most crucial role: in the context
\[  \Gamma,y::A\mid x\colon A^-,z\colon A, u\colon\hom(x,\pair+ y), v\colon \hom(\pair- y, z) \]
the variable $x$ \textit{cannot} be coerced into a term of type $A$ and the variable $z$ \textit{cannot} be coerced into a term of type $A^-$: they are not terms in the context $\Gamma,y::A\mid\bullet$, so the Neumann-Altenkirch minus operator cannot be applied, and they are not variables of type $A^\flat$, so North's $\pair+$ and $\pair-$ operators cannot be used either. For this reason, the putative motives $M(u,v)$ for the above \enQuote{induction}, either
\[  M(u,v):=\hom(\pair- y,\pair+ x) \qquad\text{or}\qquad M(u,v):=\hom(\pair- z,\pair+ y) \]
are not well-formed. This attempt at a proof of symmetry fails, as all must, since symmetry is indeed independent of this theory.

The motives $M(u,v)$ for the failed proof of symmetry in the previous paragraph are invalid, \textit{unless the type $A$ we're operating with is a core type}. That is, if $A\jEqual B^\flat$ for some $B$, then we \textit{do} have the judgment
\[ \Gamma,y::A\mid z\colon A, v\colon\hom[A](\pair- y, z)\vdash \hom[A](\pair- z, \pair+ y) \isType  \]
and so the induction ($\J[+]$) from $\refl_y\colon\hom[A](\pair- y,\pair+ y)$\footnote{Note we make liberal use of \textsc{Core-Negation} and \textsc{Core-Idem} to make this well-formed.} carries through to give us symmetry. So the hom-types for core types \textit{are} symmetric. Accordingly, we adopt the notation $\Id[A](s,t)$ instead of $\hom[A](s,t)$ when $A$ is of this form.

As in the case of \textit{un}directed type theory~\cite{van2011types}, iterating our hom-types allows us to write down higher and higher category-theoretic structure. \textit{A priori}, our types do not just carry the structure of synthetic categories, but synthetic $\infty$-categories! We leave it to future work to determine how well this framework does at capturing higher categories, but for the present work we want to restrict ourselves to lower dimensions. Neumann and Altenkirch introduce laws to restrict their their theory to synthetic (1,1)-categories (obtaining a \enQuote{(1,1)-directed type theory}), but we're more interested in the preorder case, that is, synthetic \textit{(0,1)-categories}.  So we need to make this restriction syntactically. This is done in \cref{tag:11catRules}. The \textsc{Hom Neutrality} rule asserts that every hom-type is \enQuote{neutral} in the sense of being its own core. Thus the hom-types \textit{of hom-types} are actually identity types. The \textsc{UHP} rule---the \textit{uniqueness of hom-terms principle}---asserts that there is at most one hom-term between any two points (up to propositional equality). Together, these assert that the hom-types of any type are a reflexive, transitive \textit{relation} on the type, a synthetic (0,1)-category, a synthetic preorder.

\begin{figure}
\begin{mathpar}
\inferrule[Hom Neutrality]{\Gamma\mid\Delta\vdash A\isType}{\Gamma\mid\Delta,x\colon A^-,z\colon A\vdash \hom(x,z)^\flat \jEqual \hom(x,z)}
\and \inferrule[UHP]{\Gamma\mid\Delta\vdash A\isType}{\Gamma\mid\Delta,x\colon A^-,z\colon A,p\;p'\colon\hom(x,z)\vdash \cc{UHP}(p,p')\colon\Id(p,p')}

\end{mathpar}
\caption{Restriction to synthetic (0,1)-categories }\label{tag:11catRules}

\end{figure}

Finally, we consider: what are the hom-terms of types of the form $A^-$ and $A^\flat$? Our intention is that $A^-$ and $A^\flat$ are the opposite preorder and the core setoid of the preorder $A$, respectively. This is what we assert in \cref{tag:opCoreHom}: that the type $\hom[A](x,z)$ is definitionally the same as $\hom[A^-](z,x)$---note that this is well-formed (we do not need to coerce $z$ and $x$ into different polarities as before) because $z\colon A$ and $A\jEqual (A^-)^-$, so $z$ is of the right form to be the domain of a hom in $A^-$ (and similarly for $x$). The \textsc{Core Hom} asserts that we can decompose a hom-term in the core type into hom-terms in either direction. Note that \textsc{UHP} saves us from having to assert that $\cc{fwd}(I)$ and $\cc{back}(I)$ compose to the identities, which we'd have to do in the 1-category case (to assert that $A^\flat$ is the core \textit{groupoid} of the \textit{category} $A$).

\begin{figure}
\begin{mathpar}
\inferrule[Opposite Hom]{\Gamma\mid\Delta\vdash A\isType}{\Gamma\mid\Delta,x\colon A^-,z\colon A\vdash \hom[A^-](z,x) \jEqual \hom(x,z)}

\inferrule[Core Hom]{\Gamma\mid\Delta\vdash A\isType}{\Gamma\mid\Delta,x\colon (A^\flat)^-,z\colon A^\flat,I\colon\hom[A^\flat](x,z)\vdash \cc{fwd}(I)\colon\hom[A](\pair+ x,\pair+ z)\\
\Gamma\mid\Delta,x\colon (A^\flat)^-,z\colon A^\flat,I\colon\hom[A^\flat](x,z)\vdash \cc{back}(I)\colon\hom[A](\pair- z,\pair- x)}

\end{mathpar}
\caption{Rules for opposite and core hom-terms. For the coercions of $x$ in \textsc{Core Hom}, note that we use \textsc{Core Negation} to view $x$ as a term of type $(A^-)^\flat$. So then $\pair+x\colon A^-$ and $\pair- x\colon (A^-)^-$, i.e. $A$. }\label{tag:opCoreHom}

\end{figure}

But notice that, for \enQuote{polar-closed} terms, i.e. $\Gamma\mid\bullet\vdash t\colon A$, this allows us to talk about the identity types \textit{induced} by the hom-types. \textit{A priori}, we do not have identity types between arbitrary terms: for $s\colon A^-$ and $t\colon A$ for some arbitrary type , we cannot, in general, form an identity type $\Id(s,t)$; we reserve the notation $\Id$ for core types, whose hom-types are provably symmetric by the above. However, now we \textit{can} induce an identity type
\[  \Id^\flat(s,t) := \Id[A^\flat](\pair{+\flat}s, \pair{+\flat}t).  \]
In words: we can form the identity type between $s$ and $t$, polar-closed terms of some arbitrary type $A$, by coercing $s$ and $t$ to be terms of $(A^\flat)^-$ and , respectively (note we use $\pair{+\flat}$ on $s$, so it ends up in $(A^\flat)^-$ rather than just $A^\flat$), which does have identity types. The identity type $\Id^\flat$ on $A$ gives us the core setoid of the preorder $\hom[A]$, the greatest equivalence relation contained in $\hom[A]$. Though we do not develop it explicitly here, note that we could have the dual modality to $\flat$, denoted $\sharp$, also with an induced identity type $\Id^\sharp$. The identity $\Id^\sharp$ would instead be the least equivalence relation containing $\hom[A]$, i.e. its equivalence-relation closure: from either $\hom(s,t)$ or $\hom(-t,-s)$, obtain $\Id^\sharp(s,t)$. These induced identity types will prove interesting to consider in \cref{syntheticRewriting}.

\begin{figure}
\begin{mathpar}
\inferrule[Hom Elim$+$]{
    \Gamma,y::A\mid z\colon A, v\colon\hom(\pair- y, z)\vdash M(v)\isType \\
    \Gamma,y::A\mid\bullet \vdash m\colon M(\refl_y)
}
{\Gamma,y::A\mid z\colon A, v\colon\hom(\pair- y, z)\vdash \J[+]\;m\colon M(v)}
\and
\inferrule[Hom Elim$-$]{
    \Gamma,y::A\mid x\colon A^-, u\colon\hom(x,\pair+ y) \vdash M(u)\isType \\
    \Gamma,y::A\mid\bullet \vdash m\colon M(\refl_y)
}
{\Gamma,y::A\mid x\colon A^-, u\colon\hom(x,\pair+ y)\vdash \J[-]\;m\colon M(u)}
\end{mathpar}
\caption{One-sided J-rules}\label{tag:oneSidedJRules}

\end{figure}

\section{Semantics}\label{semantics}
    %% ./semantics.tex
%%
%% By Jacob Neumann (jacobneu.com)
%% October 2025
%%

We now turn our attention to semantics. Our first step will be to expound \textit{a} model of our formal theory. Having at least one model is useful for developing a language because (a) it provides a way to check any putative rules to quickly ensure that they won't introduce disastrous consequences, and (b) it provides a mechanism to reason metatheoretically about the language, e.g. to establish independence or conservativity results. As mentioned in the introduction, these are particularly relevant tasks for judgmental directed type theorists---with the specter of symmetry ever hanging over us, we need tools to reason about what our formal language \textit{can} and \textit{cannot} do.

But we also aspire towards a more general-purpose model theory. That is, we do not just want to make singular \textit{ad-hoc} interpretations of our formal language, but wish to articulate an abstract notion of \enquote{model}, reason about constructions and structure-preserving maps between models, and so on. To our knowledge, no abstract notion of \enquote{model} has been given for crisp type theory (e.g. Licata et al.~\cite{licata2018internal} only focus on presheaf models) nor is there a common model notion for dual-context type theories, analogous to how \textit{categories with families} provide a common base which can be extended to model any structural type theory. These considerations are too involved to fully address here, but we include some reflections on model notions relevant to the present theory which, we hope, will serve as a starting-point for deeper investigations.

\subsection{The Groupoid-Category and Setoid-Preorder model}

To begin, we call to mind the notion of a \textit{category with families (CwF)}---originally due to Dybjer~\cite{dybjer}---which will serve as the foundation for the rest of our investigation.
\mkDefn{CwF}
A CwF provides the structural \enQuote{skeleton} of a model of type theory; it encodes the basic dynamics of contexts, types, substitutions, variables, etc., but is otherwise empty: to articulate the appropriate model notion for any given type-theoretic language, one must supplement the definition of \enquote{CwF} with type- and term-constructors. For our purposes, we're mostly interested in purely structural matters, and will leave consideration of specific type- and term-formers to other work.

The natural place to look for semantics for directed type theory is in the \textit{category of (small) categories}. The split between the present theory and simplicial type theory can be understood as two responses to the difficulties inherent to such a project: one finds quite quickly that the issues of \textit{variance} are quite vexing when trying to model type theory in categories. Simplicial type theory sidesteps these issues by moving to simplicial sets as their model theory instead of categories; our type theory is the result of remaining in \textbf{Cat} and developing the elaborate polarity calculus necessary to make direct category-theoretic semantics work. Accordingly, the following \textit{category model} is the paradigmatic model of this style of type theory.

\mkDefn{familySection}

\mkDefn{categoryModel}
The category model\footnote{First defined by North, though not referred to as such by her.} generalizes the \textit{groupoid model} of Hofmann and Streicher~\cite{groupoidModel}. Indeed, we get several models by restricting the above definition:
\begin{itemize}
\item if we insist every context $\Gamma$ is a \textit{groupoid} and every type $\set{A_\gamma}_{\gamma\in\Gamma}$ is a family of \textit{groupoids} (i.e. replace every instance of \enquote{category} with \enquote{groupoid} in \defnRef{categoryModel}), we obtain the \keyword{groupoid model of type theory}~\cite{groupoidModel};
\item if we insist every $\Gamma$ and every $A_\gamma$ has subsingleton hom-sets (i.e. a \textit{preorder}),\footnote{Every preorder $(X,\leq)$ can be viewed as a category whose objects are the elements of $X$ and where the hom-set $X(x_0,x_1)$ is $\set{\star}$ if $x_0\leq x_1$ and $\emptyset$ otherwise.} we obtain the \keyword{preorder model of type theory}~\cite[Chapter 2]{neumann2025ageneralized};
\item if we insist every $\Gamma$ and every $A_\gamma$ is a groupoid \textit{and} has subsingleton hom-sets (i.e. a \textit{setoid}, a set equipped with an equivalence relation), we obtain the \keyword{setoid model of type theory}~\cite{altenkirch1999extensional}.\footnote{Hofmann's \enQuote{setoid model}~\cite{hofmann1995extensional} uses \textit{partial} equivalence relations instead of Altenkirch's definition, which takes \enQuote{setoid} to mean a set with a (total) equivalence relation. Whenever we say \enQuote{setoid model}, we mean in Altenkirch's sense.}
\end{itemize}
In the present work, we're interested in a 0-truncated directed type theory (naturally modeled by the \textit{preorder model}) and its \enQuote{neutral} core (the \textit{setoid model}). However, it is actually easier to write out all our definitions for the groupoid and category models, and then later to \enQuote{truncate} away the higher-dimensional structure. So, in what follows, we'll make our definitions for categories and groupoids, and the setoid-preorder constructions can be obtained simply by insisting that all the hom-sets involved are subsingletons.

In the type theory we laid out in the previous section, each context has two zones: neutral and polarized. As the reader has likely guessed by this point, we'll interpret the former by groupoids and the latter by categories. That is, we'll articulate the \keyword{groupoid-category model} to interpret our dual-context type theory. Recalling the rules of \cref{tag:polarityCalcRules}, we begin as follows.
\begin{itemize}
\item A context $\Gamma\mid\Delta$ consists of a groupoid $\Gamma$, and a family of categories $\set{\Delta_\gamma}_{\gamma\in\Gamma}$ over $\Gamma$.
\item A type in $\Gamma\mid\Delta$ consists of a \enQuote{dependent family of categories}, i.e. a family of categories
\[ \set{A_{\gamma,\delta}}_{\gamma\in\Gamma, \delta\in\Delta_\gamma} \]
along with a functor $A_{\gamma_{01},\delta_{01}}\colon A_{\gamma_0,\delta_0} \to A_{\gamma_1,\delta_1}$ for each morphism $\gamma_{01}\in\Gamma(\gamma_0,\gamma_1)$ and $\delta_{01}\in \Delta_{\gamma_1}(\Delta_{\gamma_{01}}\;\delta_0,\delta_1)$, satisfying the appropriate functoriality laws.
\item A term $t$ of type $A$ is a section, that is, a functorial assignment
\begin{align*}
(\gamma,\delta)\qquad&\mapsto\qquad t(\gamma,\delta)\in A_{\gamma,\delta} \\ (\gamma_{01},\delta_{01})\qquad&\mapsto\qquad t(\gamma_{01},\delta_{01})\in A_{\gamma_1,\delta_1}(A_{\gamma_{01},\delta_{01}}\;t(\gamma_0,\delta_0),\,t(\gamma_1,\delta_1) )
\end{align*}
\item Extension of the polar zone $\Delta$ by $A$ consists of taking the \enQuote{total category} of $A_{\gamma,-}$ as a family of categories over $\Delta_\gamma$, for each $\gamma\in\Gamma$: the category $(\Delta\cExtend A)_\gamma$ has pairs $(\delta\in\Delta_\gamma, a\in A_{\gamma,\delta})$ as objects and morphisms $(\delta_{01}\in\Delta_\gamma(\delta_0,\delta_1), a_{01}\in A_{\gamma,\delta_1}(A_{\iden_\gamma,\delta_{01}}\;a_0,a_1))$; the functor $(\Delta\cExtend A)_{\gamma_{01}}$ is defined using the morphism parts of $\Delta$ and $A$ in the appropriate way. The first and second projection operations supply a functor $(\Gamma\mid \Delta\cExtend A) \to (\Gamma\mid \Delta)$ and a \enQuote{variable} term of type $A$ (weakened into $\Gamma\mid\Delta\cExtend A$), respectively.
\item The operation of \textit{type negation} is interpreted by applying the \textit{opposite category} operation fiberwise: $(A^-)_{\gamma,\delta} := (A_{\gamma,\delta})\op$ and viewing each functor $A_{\gamma_{01},\delta_{01}}\colon A_{\gamma_0,\delta_0}\to A_{\gamma_1,\delta_1}$ as a functor $A_{\gamma_0,\delta_0}\op\to A_{\gamma_1,\delta_1}\op$. This operation is definitionally involutive, as required by \textsc{Negation Involution}.
\end{itemize}
So, thus far, things work about the same as in the ordinary category model, just with an extra layer of dependency. The key part comes into play when we interpret the core/neutral aspect of the theory.
As is well-known in category theory, the inclusion $\mathbf{Grpd}\hookrightarrow\mathbf{Cat}$ has a right adjoint: the \textit{core groupoid} operation, which sends each category $\mc C$ to its maximal subgroupoid, the groupoid $\core{\mc C}$ with all the same objects as $\mc C$ but only the $\mc C$-\textit{iso}morphisms as morphisms. Applying this fiberwise, we get an interpretation for the $(\underlines)^\flat$ operation: $(A^\flat)_{\gamma,\delta}$ is just $\core{(A_{\gamma,\delta})}$, regarded as a category. So the morphism part of a term $e\colon A^\flat$ at $\gamma_{01}\in\Gamma(\gamma_0,\gamma_1)$ and $\delta_{01}\in\Delta_{\gamma_1}(\Delta_{\gamma_{01}}\;\delta_0,\delta_1)$ is an $A_{\gamma_1,\delta_1}$-isomorphism
\[  A_{\gamma_{01},\delta_{01}}\;e(\gamma_0,\delta_0) \quad\cong\quad e(\gamma_1,\delta_1).  \]
These are the same semantics that North gives for the core operation on types. Now, the key insight of Neumann and Altenkirch was that, in a groupoid context (a context of the form $\Gamma\mid\bullet$), one can turn around terms between $A$ and $A^-$: given terms $s\colon A^-$ and $t\colon A$ in $\Gamma\mid\bullet$, define $-s\colon A$ and $-t\colon A^-$ to have the same object parts ($(-s)_\gamma:=s_\gamma$ and $(-t)_\gamma := t_\gamma)$ but use the fact that $\Gamma$ is a groupoid to invert the morphism parts:
\begin{alignat*}{2}
(-s)_{\gamma_{01}} &:= A_{\gamma_{01}}\;(s_{\gamma_{01}\inv}) &&\in A_{\gamma_1}(A_{\gamma_{01}}\;s_{\gamma_0}, s_{\gamma_1})\\
(-t)_{\gamma_{01}} &:= A_{\gamma_{01}}\;(t_{\gamma_{01}\inv}) &&\in A_{\gamma_1}(t_{\gamma_1}, A_{\gamma_{01}}\;t_{\gamma_0}).
\end{alignat*}
But observe: $(-t)_{\gamma_{01}}$ is an inverse of $t_{\gamma_{01}}$ (likewise for $s$). That is, the morphism parts of $t$ and $-t$ together constitute the morphism part of a term of $A^\flat$. This gives semantics for the term $\pair{+\flat}t$:
\begin{equation}
(\pair{+\flat}t)_{\gamma}:=t_{\gamma} \qquad (\pair{+\flat}t)_{\gamma_{01}}:= t_{\gamma_{01}} \qquad ((\pair{+\flat}t)_{\gamma_{01}})\inv := (-t)_{\gamma_{01}}
\end{equation}
and similarly for defining $\pair{-\flat}s\colon A^\flat$. The key point to recognize is that this would not work were $\Delta$ nonempty, as we're not able to invert the morphisms of $\Delta_\gamma$.

The semantics of neutral-zone extension are related. If $A$ is a type in $\Gamma\mid\bullet$, i.e. a family of categories over the groupoid $\Gamma$, then, in order to define the \textit{neutral} extension of $\Gamma$ by $A$--- which we might denote $\Gamma\cExtend[0] A$---then the definition given for polar extension given above doesn't suffice: if we say the objects of $\Gamma\cExtend[0] A$ are pairs $(\gamma\in\Gamma,a\in A_\gamma)$ and morphisms $(\gamma_0,a_0)$ to $(\gamma_1,a_1)$ are pairs
\[  \gamma_{01}\in\Gamma(\gamma_0,\gamma_1),\quad a_{01}\in A_{\gamma_1}(A_{\gamma_{01}}\;a_0, a_1) \]
then this doesn't define a groupoid (as we can't necessarily invert $a_{01}$). Therefore, we instead define the morphisms of $\Gamma\cExtend[0]A$ to have $a_{01}$ an \textit{isomorphism}, so we get a groupoid. This is why the variables $y::A$ come out as terms of $A^\flat$ in the \textsc{Core Variable} rule: their morphism parts are isos to begin with.

That's all we'll say for the specific groupoid-category (and setoid-preorder) model semantics here. The interpretations of the hom-types and $\refl$ are the same ones that appear in North, and our semantics for $\J[\pm]$ combines Neumann's semantics for $\J[+]$ and $\J[-]$ in the obvious way.

\subsection{Towards an abstract model notion for dual-context theories}

We now wish to begin considering a general-purpose model theory for dual-context type theories. The notion of \enquote{CwF} serves as a common base for all varieties of structural type theory: to design a notion of \enquote{model} for whatever type theory we're interested in, we can start with CwFs and just attach the relevant features. But, to our knowledge, it is not so for dual-context type theories. While, for instance, crisp type theory is introduced along side a semantics (see e.g. \cite{licata2018internal}), they only consider models of a specific form, namely presheaf models. While this does supply a variety of models (much more than the two specific models we've provided for our theory), it does not reach the level of generality of CwFs. In particular, if one wanted to discuss the \textit{term model} of a given theory and prove its initiality among models, it is necessary to abstractly define what a \enquote{model} is.

We hope to contribute to this situation. It is helpful that our dual-context theory is somewhat, but not hugely, different from crisp/cohesive/spatial type theories (for instance, our $A^\flat$ type can be formed in any context, whereas Shulman~\cite[Figure 5]{shulman2018brouwer} needs $\Gamma\mid\bullet\vdash A\isType$ in order to form his type $\flat A$); this (hopefully) illuminates what's the bare-bones structure of dual-context theories and what's specific to a given theory.

So here we give our notion of \enquote{dual CwF}. The idea with this notion is to capture just the basic situation of dual-context type theory: that there are two zones to the context, one depending on the other, and that we form types and terms depending on both. We employ our notion of \enquote{family of categories} and \enquote{section} from before (\defnRef{familySection}), but here to specify the structure possessed by an abstract model rather than to construct a specific model.
We begin by defining the auxiliary notion of the \textit{total category}, which will be familiar to those who've studied Grothendieck fibrations
\mkDefn{totalFamCat}
\mkDefn{dualCwF}

The category $\mc C$ interprets the leftmost context zone (the \enQuote{crisp} or, in our case, \enQuote{neutral} variables). It is not a CwF in its own right, as this definition does not provide it a context extension operation: the question of how to extend the leftmost zone is, we think, theory-specific. Within a given context $\Gamma$ of $\mc C$, we have a CwF of the polar/cohesive contexts $\mc D_\Gamma$: the empty context $\Gamma\mid\bullet$, the formation of types and terms ($\Gamma\mid\Delta\vdash\mc J$), and the context extension operation, which extends $\Delta$ but leaves $\Gamma$ untouched (this is what the \enQuote{first projection} condition achieves).

As an example of how this calculus works, suppose we have a $\mc C$-morphism $\sigma\in\mc C(\Theta,\Gamma)$. Intuitively, this is a substitution which just operates on neutral/crisp variables, implementing the variables of $\Gamma$ in terms of the variables of $\Theta$. The contravariance of the family $\mc D$ says that $\sigma$ induces an operation on polar/cohesive context zones:
\[   \Delta\in \mc D_\Gamma \quad\mapsto\quad \Delta[\sigma]\in\mc D_\Theta. \]
This substitution can be regarded as a substitution in the broader CwF: observe by \defnRef{totalFamCat} that we can pair $\sigma$ with $\iden_{\Delta[\sigma]}\in\mc D_\Theta(\Delta[\sigma],\Delta[\sigma])$ to obtain a morphism in the total category
\[  (\sigma,\iden) \quad\colon \Sub\;(\Theta\mid \Delta[\sigma])\;(\Gamma\mid\Delta).  \]
Thus, types and terms in $\Gamma\mid\Delta$ can be substituted to types and terms in $\Theta\mid\Delta[\sigma]$.

We also include an extension operator for the neutral context zone, but only by types with no polar dependencies, i.e.
\[  \Gamma\in\mc C,\ A\in\Ty(\Gamma\mid\bullet) \mapsto (\Gamma\cExtend[0]A)\in\mc C.  \]
This is the kind of theory-specific operator which can be added to the basic definition of dual CwF. To define the abstract notion of \enquote{model of dual-context judgmental directed type theory}, we would need to add numerous constructors (corresponding to the rules listed in \cref{typeTheory}, among others), and likewise to define the notion of \enquote{model of crisp type theory}. But both could use dual CwFs as a starting-point, as it represents the basic structural foundation of dual-context type theory.

That's all we'll say on this matter for the present moment. Again, we do not pretend to have exhausted the matter of an abstract model theory for dual-context type theories. But we hope we have provided a starting point for deeper investigations into this topic.

\section{Application: Synthetic Rewriting Systems}\label{syntheticRewriting}
    %% ./syntheticRewriting.tex
%%
%% By Jacob Neumann (jacobneu.com)
%% October 2025
%%
Finally, we briefly consider an application of this language: a synthetic framework for reasoning about \textit{rewriting systems}. The basic idea is that the hom-types are intended to internalize metatheoretic reduction in the same way the identity types of \textit{un}directed type theory internalize metatheoretic equality. It's fitting that we employ a directed type theory which validates the uniqueness of homs principle (e.g. the directed type theory modeled by the setoid-preorder model), as we wish to discuss reduction as a \textit{mere relation}---we are not interested in the $\infty$-categorical structure provided by simplicial type theory, or even in the 1-categorical structure of Neumann's (1,1)-directed type theory.\footnote{We leave to future work to investigate possible connections to \enQuote{higher-dimensional rewriting}.} We suspect that the present system---which is \enQuote{directed intensional} in that we don't assert any kind of \textit{directed equality reflection} principle and allow our directed equality relation to be coarser than metatheoretic reduction---to have a similar relationship to a \enQuote{directed extensional} type theory (which \textit{does} assert such a principle) as undirected intensional type theory has to undirected extensional type theory (see e.g. \cite{hofmann1995extensional}). We leave these detailed technical questions, as well as an in-depth study of what can be done with our system, to future work---here we only wish to indicate how the interpretation goes.

In this setting, any type $A$ which we can write down (assume for simplicity that $A$ doesn't depend on any polarized variables, i.e. $\Gamma\mid\bullet\vdash A\isType$) has the structure of a \keyword{synthetic rewriting system}. We'll think of the terms of $A$ as \textit{expressions} and the hom-type relation as \textit{reduction}; indeed, we could write
\[  s \reduces t \]
instead of $\hom(s,t)$ to suggest this interpretation. Of course, this $\reduces$ relation is reflexive: \textsc{Hom Intro} asserts that $\Gamma,y::A\mid\bullet\vdash \refl_y\colon \pair-y \reduces \pair+y$, and, in particular, for any $\Gamma\mid\bullet\vdash t\colon A$, we have $\refl_{\pair{+\flat}t}\colon -t \reduces t$. We also know automatically that the reduction relation is transitive, by the composition of hom-terms operation provable from the rules of directed type theory.

Using the $\J[+]$ rule we can prove a transport law of the form
\begin{mathpar}
\inferrule{\Gamma\mid\bullet\vdash s\colon A^-\\
\Gamma\mid z\colon A\vdash M(z)\isType\\
\Gamma\mid\bullet\vdash m\colon M(-s)}
{\Gamma\mid z\colon A, f\colon s \reduces z\vdash \cc{tr}^+_M\;f\;m := \J[+]\;m\colon M(z)}
\end{mathpar}
Intuitively: any property $M(z)$ we can write down in this language is \textit{preserved under reduction}. That is, if $M$ holds of $s$ and $s\reduces z$, then $M$ must hold of $z$ as well. Using $\J[-]$, of course, we can prove the same claim running backwards, namely that if $M$ holds of $t$ and $x\reduces t$, then $M$ must hold of $x$ too. Now, perhaps this is \enQuote{too nice}, and we might want to write down more syntactic properties (e.g. \enQuote{is a value}) or relations among terms (e.g. \enQuote{is a subexpression}) which are not closed under reduction in this way. Perhaps some system with a further context zone, where constructions need not even be functorial in variables from that zone?

Though it is beyond our scope to precisely articulate a theory of \textit{directed quotient inductive types}, some informal use will help illustrate our point. Recall from \textit{un}directed type theory that a higher inductive type (HIT) is a type generated from a set of constructors (in the same way an ordinary inductive type is generated by constructors, e.g. $\N$ from $zero$ and $succ$), but where some of the constructors instead construct \textit{identities} of the type being constructed. This is used e.g. in homotopy type theory~\cite[Section 6]{hottbook} to construct higher-dimensional homotopy spaces. But the same technique is useful even in the set-truncated world: quotient inductive types allow one to build interesting mathematical structures (e.g. the integers or real numbers) as a quotient of an inductive type. The directed version is to introduce a type by constructors, including constructors of the hom-type. For instance, we could introduce the type $\mathbf{Expr}$ of \textit{natural number expressions} via the constructors:
\begin{itemize}
\item $\vdash 0\colon\mathbf{Expr}$
\item $a\colon \mathbf{Expr} \vdash \cc{succ}(a)\colon\mathbf{Expr}$
\item $a\;b\colon\mathbf{Expr} \vdash a+b\colon\mathbf{Expr}$
\item $a\colon\mathbf{Expr} \vdash \cc{pz}(a)\colon a + 0 \reduces a$
\item $a\;b\colon\mathbf{Expr} \vdash \cc{ps}(a,b)\colon a+\cc{succ}(b)\reduces \cc{succ}(a + b)$
\end{itemize}
(we omitted the polarity coercions here for clarity). So, combining the $\cc{pz}$ and $\cc{ps}$ constructors in the right way (and liberally using the transport operation above to reduce subexpressions), we can construct internal proofs about this simple reduction calculus. This is of course a very simple example, but serves as a proof of concept.

Combine this with our induced identity types $\Id^\flat$ and $\Id^\sharp$. Recall that we had an identity $\Id^\flat(s,t)$ when we had $s\reduces t$ and $-t\reduces -s$. Interpreting our types as synthetic rewriting systems, $\Id^\flat$ encodes some kind of syntactic, literal, intensional equality between $s$ and $t$: for a \enQuote{reasonable} notion of rewriting, we would expect that the only way two terms could rewrite to each other is if they're already literally the same. On the other hand, $\Id^\sharp$ gives us a notion of extensional equivalence: it is the equivalence relation generated by reduction, so if $s\reduces t$, or $-t\reduces -s$, or they reduce to some common expression, etc. then they are equal under $\Id^\sharp$. Again, if we make some assumptions of \enQuote{niceness} for our reduction relation (e.g. confluence), then $\Id^\sharp(s,t)$ if they reduce to a common expression. So, for our example above, we'd have a proof of $7+5\reduces 12$ and therefore $\Id^\sharp(7+5,12)$, but, since we couldn't construct a term of type $12\reduces 7+5$, we wouldn't get $\Id^\flat(7+5,12)$, as we might hope.

\bibliographystyle{alpha}
\bibliography{biblio.bib}
\end{document}